\documentclass[a4paper,twoside]{article}
\newcommand{\affil}[1]{$^{\rm #1}$}
\usepackage[authoryear]{natbib}
\bibpunct{(}{)}{;}{a}{}{,}
\usepackage{graphicx}
\date{} 
\title{\large\bf\flushleft Photometric analysis of Magellanic Cloud R Coronae Borealis Stars in the recovery phase of their declines \footnote{Based partially on observations made with the Southern African
Large Telescope
  (SALT)}}
\author{\parbox{\textwidth}{\flushleft
\vspace{-0.5cm}
{\it Robyn M. Woollands\affil{A}, P.L. Cottrell\affil{A,C}, and A. Udalski\affil{B}}\\
\vspace{0.4cm}
{\small \affil{A}\,Dept of Physics \& Astronomy, University of Canterbury, New Zealand }\\
{\small \affil{B}\,Warsaw University Observatory, Al. Ujazdowskie 4 00-478 Warszawa, Poland}\\
{\small \affil{C}\,Email: peter.cottrell@canterbury.ac.nz}}}
\begin{document}
\twocolumn[
\begin{changemargin}{.8cm}{.5cm}
\begin{minipage}{.9\textwidth}
\vspace{-1cm}
\maketitle
%
%
\small{\bf Abstract:}

This paper presents the initial results of a multi-site photometric programme to examine the extraordinary behaviour displayed by 18 R Coronae Borealis (RCB) stars in the Magellanic Clouds (MCs). RCB stars exhibit a unique variability whereby they undergo rapid declines of up to several magnitudes. These are thought to be caused by the formation of dust in the stellar environment which reduces the brightness.

The monitoring programme comprised the collection of $UBVRI$ photometric data using five telescopes located at three different southern hemisphere longitudes (Las Campanas Observatory in Chile, Mount Joun University Observatory in New Zealand, and the Southern African Large Telescope (SALT) in South Africa).

Examination of the data acquired in the $V$ and $I$ filters resulted in the identification of a total of 18 RCB declines occurring in four stars. Construction of colour-magnitude diagrams ($V$ vs $V-I$), during the recovery to maximum light were undertaken in order to study the unique colour behaviour associated with the RCB declines. The combined recovery slope for the four stars was determined to be $3.37\,\pm\,0.24$, which is similar to the value of $3.1\,\pm\,0.1$ calculated for galactic RCB stars (Skuljan et al. 2003). These results may imply that the nature of the dust (i.e. the particle size) is similar in both our Galaxy and the MCs.

\medskip{\bf Keywords:} stars: variable: other --- stars: activity --- Magellanic Clouds

\medskip
\medskip
\end{minipage}
\end{changemargin}
]
\small

\section{Introduction}

R Coronae Borealis (RCB) stars are a type of eruptive variable that undergo rapid and enigmatic magnitude declines of variable amplitude. They are a rare type of 
helium- and carbon-rich, but hydrogen-poor, supergiant star with only 51 having been identified in the Galaxy (Tisserand et al. 2008) and a further 26 in the Magellanic 
Clouds (Alcock et al. 2001). If it is a common process why don't all stars show hydrogen deficiency? This implies that these hydrogen-deficient objects must be produced by an unusual combination of stellar nucleosynthesis and other processes. Two scenarios have been proposed as the most likely. These are the merger of two types of white dwarf stars in a binary system or a brief nuclear burning phase that reinflates the object (so-called ñborn againî objects) back into a supergiant star that is more easily observable to astronomers (see Clayton 1996).

\begin{figure}[h]
\begin{center}
\includegraphics[scale=1, angle=0, width = 80mm]{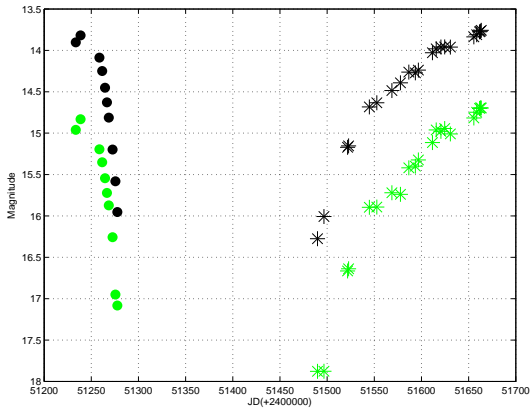}
\caption{A typical RCB light curve (LMC104.4 60857), based on data presented in this paper, displaying a characteristic decline (dots) and recovery (stars) associated with the obscuration of the photosphere by the formation of dust along the line-of-sight. The green and black points represent the observations taken in the $V$ and $I$ filters respectively.}\label{TypicalDec}
\end{center}
\end{figure}

The random variabilty displayed by RCB stars is a rapid decline in the magnitude followed by a gradual recovery (Figure ~\ref{TypicalDec}). A decline usually occurs over a time-span of days or weeks and may be as much as an eight magnitude drop below maximum brightness (Clayton 1996), whereas the recovery to normal brightness may take months to years. The depth, shape and duration of a decline may differ significantly from one star to another and from one decline to the next. It is generally thought that the 
declines are the result of obscuration of the photosphere by the presence of newly formed amorphous dust along the line-of-sight that eclipses the star (Clayton 
1996). Each successive drop in magnitude within a large decline may be explained by a new dust cloud forming along the line-of-sight. The mechanism responsible for gas ejection and the formation of carbon dust, close to the photosphere, is not yet known (Walker 1985; Whitney et al. 1992). However, it is widely believed that the dust condensation is strongly related to the pulsational variability of RCB stars (Fadeyev 1983).

Distances to the galactic RCB stars, and hence their absolute magnitudes, are currently ill-defined. However the 26 RCB stars identified in the Magellanic Clouds 
(Alcock et al. 2001) do not suffer from poor absolute magnitude determination as the distances to the Magellanic Clouds are precisely known. Therefore these stars have well-defined positions in the $L-T{_{\rm eff}}$ planes for intercomparison and linking with evolutionary tracks.  In particular, studying the RCB stars in the metal-poor environment of the MCs is important for making comparisons between 
themselves and the galactic RCB stars, through the influence that chemical composition may have on their respective declines. In addition, their mysterious evolutionary status and the fact that they produce large quantities of dust allows dust formation and evolution processes to be studied. This means that the study of 
RCB stars has important consequences in the understanding of a number of properties of the stellar and circumstellar environment of these enigmatic stars.

\section{Instruments \& Observations}

Photometric data analysed for this paper were collected using five different telescopes situated at three observatories, each positioned 
on a separate southern hemisphere continent (South America, Australasia and Africa). 

At Las Campanas Observatory in Chile, $V$ and $I$ data were 
collected using the 1.3\,m Warsaw telescope in the course of the long term sky survey --
the Optical Gravitational Lensing Experiment (OGLE). At Mount John 
University Observatory (MJUO) in New Zealand, $BVRI$ photometric data were acquired with the Optical Craftsmen (OC) and the Boller and Chivens (B\&C) 
0.6\,m telescopes. The instruments used in conjunction with the OC and B\&C telescopes were respectively the SBIG STe9 CCD system and filter wheel (accommodating the Schuler 
$UBVRI$ filters in the Johnson/Cousins photometric system) and the Apogee Alta CCD system (same set of filters). The remaining 
data were collected in the $UBVRI$ filters using the 11\,m Southern African Large Telescope (SALT) and SALTICAM, in South Africa.

The OGLE data were collected over two intervals. The first spanned January 1997 to November 2000 (OGLE-II), and the second July 2001 to February 2008 (OGLE-III). Four observing runs (two using the OC and two using the B\&C) were undertaken at MJUO. The two OC observing runs lasted about one 
and a half weeks and occurred in mid September and early October, 2007. The second OC observing run occurred simulatneously with the first B\&C observing run. This was 
essential in order to allow accurate photometric calibration of the data obtained from these telescopes (see Section ~\ref{RedCal}). The final B\&C observing run occurred 
in January 2008. SALT observations were collected over a time lasting from October 2007 to February 2008. All in all, a total of about $3000$ observations of the 18 RCB stars in the Large and Small Magellanic Clouds (LMC and SMC) were collected. 

\section{Reduction \& calibration}
\label{RedCal}

Data for this paper were collected over the previous $11$ years by four different telescopes, and as a result, slightly different procedures were followed during the preparation of different datasets prior to their reduction. The MJUO images required the 
most pre-processing while the OGLE data were provided as calibrated $V$ and $I$ magnitudes.

To determine the magnitudes of the target stars in the MJUO and SALT frames, they were reduced using the aperture photometry methods within
Mira AL \footnote[1]{The Mira AL software was developed as a tool for image processing with advice from affiliates of the Hands On Universe (HOU) project (Lawrence Hall of Science, University of California, Berkeley), and the Conceptual Astronomy and Physics Education Research (CAPER) project (Department of Astronomy, University of Arizona).}. This process involves comparing the counts obtained from the standard stars in the field (with known magnitudes), to the counts obtained 
from the target star. At least two, and in most cases three, standard stars were present in each field. Each star in the field that is selected for measuring needs three apertures. The central one measures the combined signal from the star 
and the sky. The middle and outer apertures form an annulus within which only the signal from the local sky is measured. Subtracting this from 
the signal obtained from the inner aperture (combined star and sky) gives the signal detected from the star alone. Mira AL computes the magnitude 
using this value, the various FITS header keywords (GAIN, EXPTIME, ZERO-POINT) and the photometric zero point from the standard stars in the field.

Point spread function (PSF) methods were not used in this research as the stars were chosen to be in relatively isolated fields and there were significant image quality variations across the field of the SALTICAM data and from one frame to the next. This meant that the PSF was variable in time and position in the CCD frames making measurement and application of this technique extremely difficult. 

In order to reduce the data in a way that will produce precision measurements, particular attention must be paid to the following: the sky annulus 
should be positioned strategically so as to avoid the inclusion of other stars and irregularities in the background that would not otherwise 
exist underneath the object; the inner aperture should not be too large that it extends to the place where the star appears to merge into the sky 
noise; and the same aperture size should be used for measuring every object.

As alluded to earlier, all the OGLE data were provided as calibrated magnitudes. The OGLE data reduction is performed using DIA (difference image
analysis) technique - for details the reader is referred to Udalski et al. (2008).

To be able to treat all the data as one dataset and make precise comparisons, data calibration between the five telescopes is essential. To do this all 
the MJUO data were calibrated using F region standard star observations that were obtained during various observing times. Observations of F region 
standard star (F116)  were taken concurrently on the OC and B\&C telescopes. Since the magnitudes of F region stars are known and constant (Cousins and Stoy 1962), it is possible to calculate the magnitude of the target star with respect to them. The F116 stellar images taken from different telescopes can be used to align all the images with a common zero-point which in turn allows the different target magnitudes to be compared and treated as one dataset. Images of F116 were also reduced using the method of aperture photometry and a 
SIMBAD identified standard star. The F region stellar magnitudes obtained from the data collected on the OC and B\&C telescopes differed slightly 
from the published values. To eliminate the inconsistencies the average magnitude in each filter for each telescope was calculated. The difference between these values and the published value in each filter for F116 was then applied to correct the data. These same shifts were then applied to all the data obtained on the OC and B\&C respectively, thus ensuring their common zero point.

Since observations of an F region standard star were not taken with SALT, an alternative approach to data calibration had to be adopted. During the 
data reduction process the magnitude of each standard star in a given field is entered manually. These magnitudes are those reported by 
SIMBAD and once they have all been recorded by Mira AL, they are automatically varied slightly with respect to the brightness and the assigned magnitudes 
of the standard stars in the field. To calibrate the SALT data the offset between the magnitudes of the recently calibrated standard stars in the MJUO 
images, and the same standard stars in the SALT images, is calculated and applied. This operation yields both the data from SALT and MJUO with a 
common zero point.

As previously mentioned, the OGLE data is received in a post-reduction form as stellar magnitudes. The only preparation that it requires are the $V$ and $I$ colour corrections. To calibrate the SALT and MJUO data with these OGLE magnitudes, several processed OGLE images were analysed in a similar manner to that described earlier. The magnitudes obtained through this reduction process were compared with the originally received OGLE magnitudes. An average offset was calculated and this was applied to all the calibrated SALT and MJUO magnitudes that already shared a common zero-point. This allowed all the magnitudes analysed to become aligned with the colour corrected OGLE data zero-point. Of course, since OGLE only provides data in the $V$ and $I$ filters, this also limited the calibration of the SALT and MJUO data to these filters. Nonetheless, calibration across the filters is certainly an issue that will be given more attention in future anaylses.

\section{Recovery phase photometry}

\begin{figure}[h]
\begin{center}
\includegraphics[scale=1, angle=0, width = 80mm]{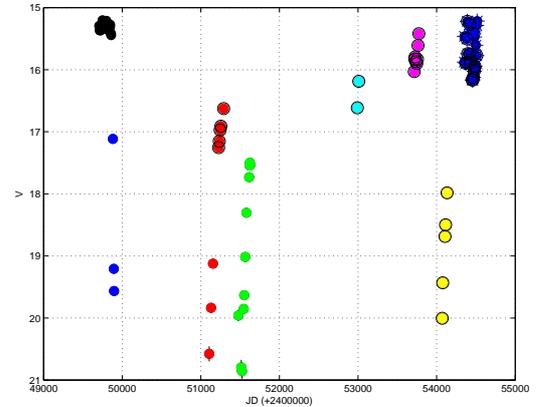}
\caption{The light curve ($V$) for
  LMC100.6 48589 displaying the seven declines (separately
  colour-coded) in the time interval of this research. The solid black points are data obtained at maximum light, while the black circled points are the data selected for use in further decline recovery calculations.}\label{LMC1006_ALL}
\end{center}
\end{figure}

\begin{figure}[h]
\begin{center}
\includegraphics[scale=1, angle=0, width = 80mm]{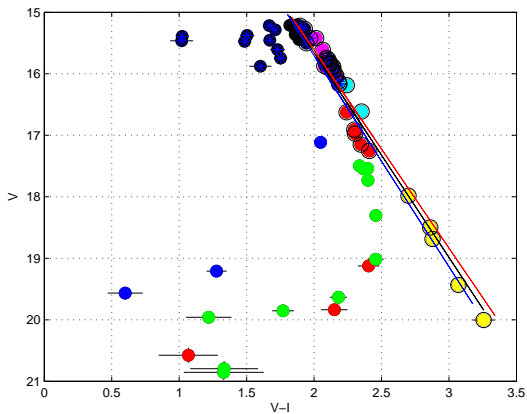}
\caption{CMD for LMC100.6 48589 showing the characteristic `loop'
  decline and `asymptotic' recovery track traced by each of the seven
  declines (colour-coded) observed in Figure~\ref{LMC1006_ALL}. The
  black line ($\frac{\triangle V}{\triangle(V-I)}=3.33\pm0.10$) represents the asymptotic recovery path and the red
  and blue lines are used in the calculation of its uncertainty.}\label{LMC1006_VI}
\end{center}
\end{figure}

Figure~\ref{LMC1006_ALL} shows the seven declines that have occurred in LMC100.6 48589 during the $\sim$ 6,000 days over which the data were collected. 
Even though seven declines have been identified only certain data points (circled in black) in a decline were selected for inclusion in 
further recovery phase calculations. This is because during the initial decline phase the dust in the stellar environment is potentially inhomogeneously distributed 
and results in the rapid drop observed in magnitude. There are also strong chromospheric spectral features visible at the early decline phase that do not decay until the recovery phase (Cottrell et al. 1990). And in the recovery phase the dust around the star is assumed to be more uniformly 
distributed, and it is at this time that a noticeable excess in the ($U$-$V$), ($B$-$V$), ($V$-$R$) and ($V$-$I$) colours becomes apparent. Since the 
late recovery phase is the desired component of the decline under investigation in this paper, a magnitude threshold was set in order 
to reject all data that were collected prior to dust homogeneity over the stellar disc. This threshold varied between individual stars and 
was also dependent on the depth to which a particular decline fell. It was generally taken as being two or three magnitudes below maximum 
brightness, which corresponds to the data points in the colour-magnitude diagram (CMD) following an asymptotic recovery line back to maximum light (Figure~\ref{LMC1006_VI}).

Figure~\ref{LMC1006_VI} shows the data obtained for LMC100.6 48589 and plotted in a CMD. It reveals that during the onset of a decline the sudden decrease in 
magnitude is portrayed by the points on the CMD moving to fainter $V$ and ultimately redder ($V$-$I$). For some declines the ($V$-$I$) colour does show 
a `blue-type' decline (Cottrell et al. 1990), with the ($V$-$I$) becoming bluer than at maximum light. However other 
declines are of the `red-type'.  As the star recovers to 
normal brightness the data points follow a characteristic `loop' track towards the asymptotic recovery line (black line). 
Skuljan et al. (2003) showed that in a sample of nine galactic RCB stars observed to have a total of 26 declines over a period of 12 years this asymptotic recovery back to maximum light had a slope of [$\frac{\Delta V}{\Delta(V-I)}$]$_{galactic}=3.1\,\pm\,0.1$. In addition, the slope 
of the asymptotic line does not depend on the depth or the frequency of declines and is the same for all these galactic RCB stars. At this phase in the recovery the various filter band-passes are no longer affected by the emission lines that are present to a greater (`blue') or lesser (`red') extent and the process is one of dispersion of the dust cloud (see Cottrell et al 1990). The determination of the asymptotic recovery slope for the 18 declines of the four MC RCB stars studied in this research is determined using this same assumption (see Table~\ref{DecData}), even though no detailed decline phase spectroscopic observations have yet been made to confirm this in these Magellanic Cloud objects.

The position of the asymptote (black line) for all figures in this paper was determined by fitting a straight line to the data 
points circled in black, and in some cases (e.g. LMC100.6 48589 in Figure~\ref{LMC1006_VI}) the points that were identified as representing the star at maximum 
light (black points). The reason for including the latter points in the calculation is that the asymptotic line must pass through them since the 
recovering star will eventually end up with a maximum magnitude that will be at this position in the CMD. The inclusion of these points is essential 
to ensure the best possible calculation for the gradient of the asymptote. The gradient of this line for LMC100.6 48589 was calculated as having a 
value of $\frac{\Delta V}{\Delta(V-I)}=3.33\,\pm\,0.10$, with the uncertainty computed using the red and blue lines in Figure~\ref{LMC1006_VI}. These are lines fitted to the extreme maximum 
and minimum possible values (taking into account their uncertainty in both $V$ and ($V$-$I$)) for the data points respectively. This method for determining 
the uncertainty was chosen because it not only takes into account the positional spread of the desired data, but it also includes their individual uncertainties. 

\begin{figure}[h]
\begin{center}
\includegraphics[scale=1, angle=0, width = 80mm]{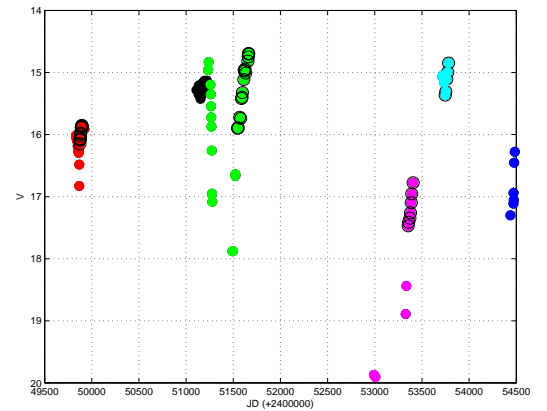}
\caption{The light curve for LMC104.4 60857 revealing declines reaching
various depths.}\label{LMC1044_ALL}
\end{center}
\end{figure}

\begin{figure}[h]
\begin{center}
\includegraphics[scale=1, angle=0, width = 80mm]{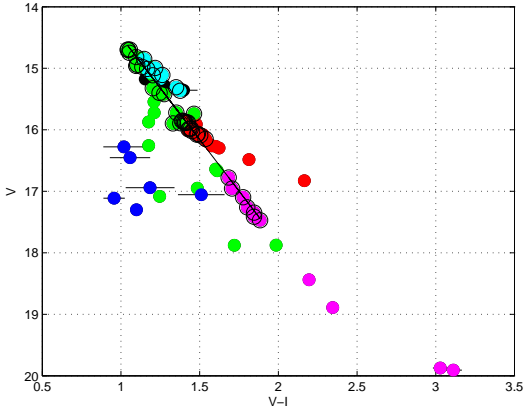}
\caption{The CMD for LMC104.4 60857.}\label{LMC1044_VI}
\end{center}
\end{figure}

Figure~\ref{LMC1044_ALL} displays the LMC104.4 60857 data. It shows several declines of varying depth. When these data are plotted in a CMD (Figure~\ref{LMC1044_VI}), 
it is evident that 
initially each decline takes a slightly different `loop' (a `red-type' according to Cottrell et al. 1990) path, but remarkably during the recovery phase all the decline data track along the same line. This provides more evidence to consolidate the 
idea that the depth of an individual decline is unrelated to the final approach to maximum light. The slope of the recovery asymptote for LMC104.4 60857 
was determined to be ($3.34\,\pm\,0.02$), the same, within the uncertainties to that calculated for LMC100.6 48589.

\begin{figure}[h]
\begin{center}
\includegraphics[scale=1, angle=0, width = 80mm]{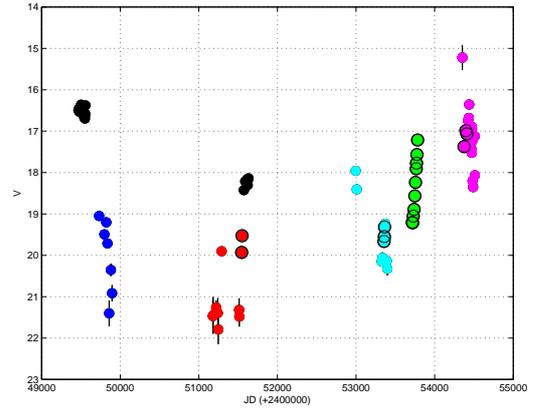}
\caption{The light curve for LMC169.7 40649 shows that it is probable
  that the star did not return to maximum brightness during the
  time-span over which the data were collected.}\label{LMC1697_ALL}
\end{center}
\end{figure}

\begin{figure}[h]
\begin{center}
\includegraphics[scale=1, angle=0, width = 80mm]{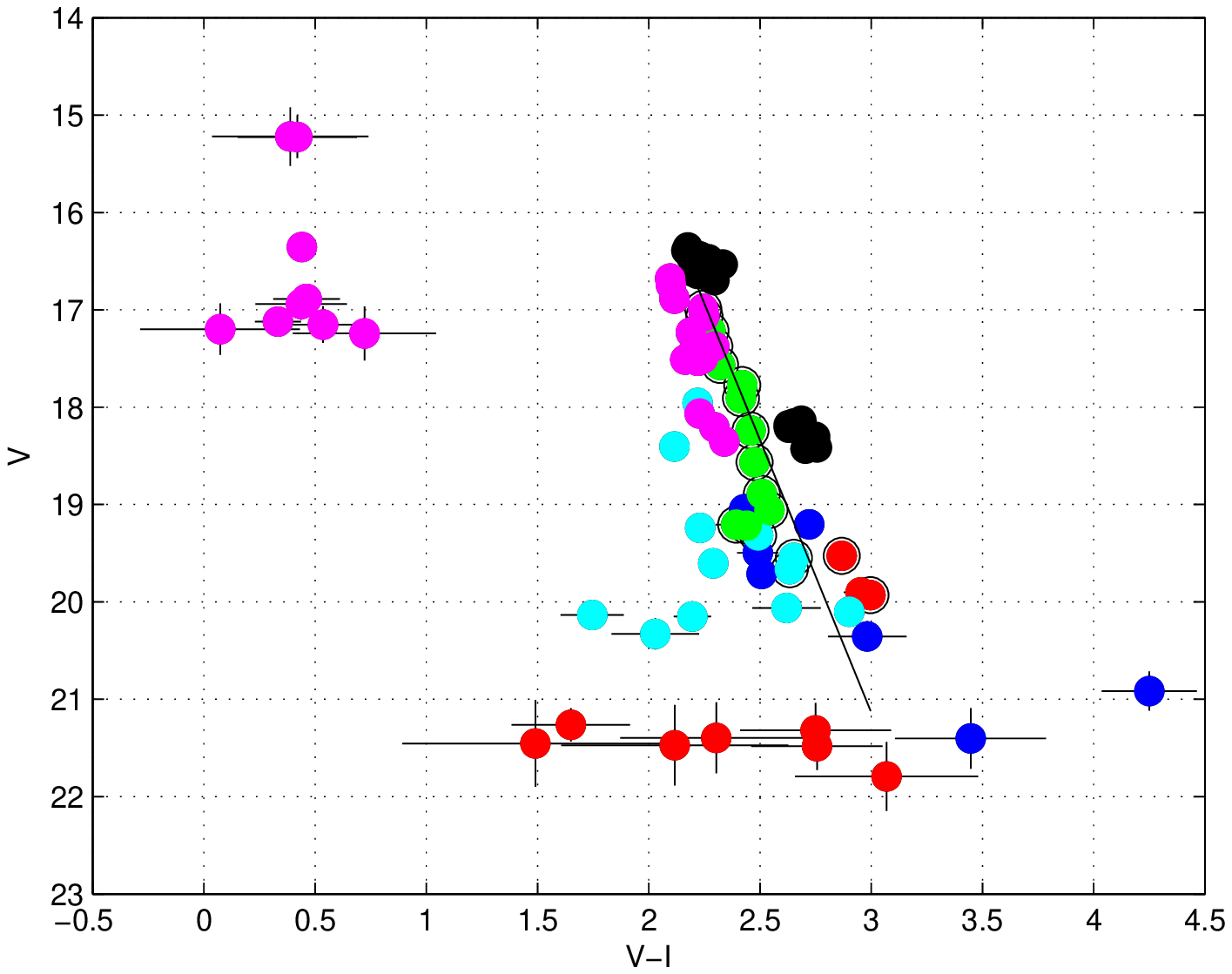}
\caption{The CMD for LMC169.7 40649 revealing a steep slope of $\frac{\triangle V}{\triangle(V-I)}=5.59\pm0.24$.}\label{LMC1697_VI}
\end{center}
\end{figure}

\begin{figure}[h]
\begin{center}
\includegraphics[scale=1, angle=0, width = 80mm]{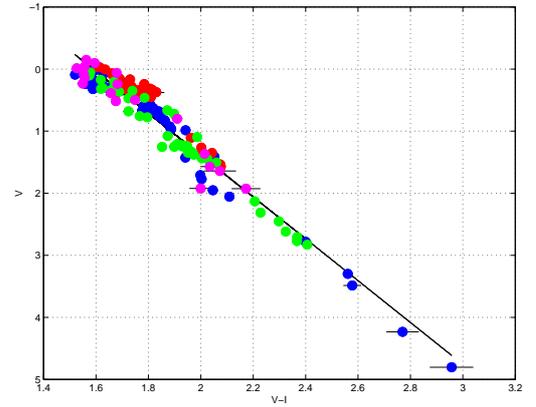}
\caption{Declines from four RCB stars with a combined gradient of
  $\frac{\triangle V}{\triangle(V-I)}=3.37\pm0.24$. The data from each
star are colour-coded as follows: LMC100.6 48589: blue, LMC101.7 8114:
red, LMC104.4 60857: green, and SMC105.6 22488: magenta.}\label{AveDec}
\end{center}
\end{figure}

Figures~\ref{LMC1697_ALL} and~\ref{LMC1697_VI} show the light curve and CMD for LMC169.7 40649. Although declines are evident in the light curve, 
it looks as if the star does not 
return completly back to maximum brightness during the time-span over which the data were collected. Instead, the data portray the star as being continuously in a 
decline, which probably accounts for the steep slope of $\frac{\Delta V}{\Delta(V-I)}=5.59\,\pm\,0.24$ observed in the CMD. As a result, the data from this star were not 
included in the combined MC slope calculation (see Figure~\ref{AveDec}).

In total 18 declines occurred in four stars over the time-span of about eight years. The mean value for the gradient of the recovery slope for these four stars is 
$\frac{\Delta V}{\Delta(V-I)}$=$3.31\,\pm\,0.10$ (Table~\ref{DecData}). To calculate a more representative recovery track gradient, the data points from the 18 
declines were shifted so that they share a common zero point. The shift was achieved by subtracting the average magnitude at maximum light from the photometric 
data (in both the $V$ and $I$ filters) to create new $V$ and ($V$-$I$) values. Following this, a straight line 
was fitted to the recovery data from these four stars in order 
to determine the combined gradient. Figure~\ref{AveDec} shows the data after the maximum magnitude subtraction was performed. The value for the gradient calculated 
using this method is [$\frac{\Delta V}{\Delta(V-I)}$]$_{combined}=3.37\,\pm\,0.24$, with the uncertainty being determined in the same way as that described earlier. 
These two gradient values are consistent considering their uncertainties. They also fall close to the value of 
[$\frac{\Delta V}{\Delta(V-I)}$]$_{galactic}=3.1\,\pm\,0.1$ for the 26 declines of nine galactic RCB stars that was previously determined by Skuljan et al. (2003).

The closeness of these two decline recovery gradients implies that the mechanism responsible for the observed magnitude declines is common to RCB stars in both the Galaxy and the Magellanic Clouds. Given that at this phase all the emission lines in the galactic RCB stars have disappeared from the spectra, one is simply looking at the scattering of light by dust. The similarity of the gradients then implies  that the size of the dust particles in the stellar environment would be similar in the Galaxy and the Magellanic Clouds. 

\begin{table}[h]
\begin{center}
\caption{The line slopes obtained from the ($V-I$) vs $V$ CMDs for four RCB
stars in the MCs.}\label{DecData}
\begin{tabular}{lcc}
\hline Star            & No. of declines & $\frac{\triangle V}{\triangle(V-I)}$ \\
\hline LMC100.6 48589 & $7$ & $3.33\pm0.10$ \\
\hline LMC101.7 8114$^*$ & $5$ & $3.34\pm0.01$ \\
\hline LMC104.4 60857 & $4$ & $3.34\pm0.02$ \\
\hline SMC105.6 22488 & $2$ & $3.21\pm0.22$ \\
\hline Total and Mean & $18$ & $3.31\pm0.10$ \\
\end{tabular}
\medskip\\
$^*$MJUO and SALT data were included in the analysis.\\
\end{center}
\end{table}

\section{Discussion}

The recovery track gradient value obtained for the MCs ([$\frac{\Delta V}{\Delta(V-I)}$]$_{combined}=3.37\,\pm\,0.24$) lies close to the previous value 
obtained for galactic RCB stars ([$\frac{\Delta V}{\Delta(V-I)}$]$_{galactic}=3.1\,\pm\,0.1$), and could imply that the chemical abundance/environment and hence 
the dust particle sizes in their respective locations are similar in both our Galaxy and the MCs. An interesting point to note is that the recovery asymptote 
for the MCs is positioned at larger values of ($V$-$I$) compared with that calculated for galactic RCB stars (Skuljan et al. 2003). This 
may be a result of increased interstellar dust 
existing along the line-of-sight in the direction towards the MCs compared with the RCB stars previously studied in the Galaxy. 

The gradient calculated for the 
LMC RCB stars alone (excluding SMC105.6 22488), is [$\frac{\Delta V}{\Delta(V-I)}$]$_{LMC}=3.34\,\pm\,0.21$. Comparing this value with the recovery gradient for just 
the SMC, or simply SMC105.6 22488 ($[\frac{\Delta V}{\Delta(V-I)}$]$_{SMC}=3.21\,\pm\,0.22$), it is clear that they also agree to within their associated uncertainties. 
This suggests that the physical properties of the dust forming in the RCB stellar environments in the LMC and SMC are also similar. 

\section{Conclusion}

The work completed for this paper involved the initiation of a multi-site photometric programme to examine the extraordinary behaviour displayed by 18 RCB stars in the MCs. The monitoring programme comprised the collection of $UBVRI$ photometric data using five telescopes located in Chile, New Zealand and South Africa.

Examination of the data acquired in the $V$ and $I$ filters, resulted in the identification of a total of 18 RCB declines occurring in four stars (three in the LMC and one in the SMC). Construction of colour-magnitude diagrams ( $V$ vs $V$-$I$), in particular during the recovery to maximum light, were undertaken in order to study the unique colour behaviour associated with the RCB declines as a result of the dust formation in the stellar environment. The combined recovery slope for the four stars was determined to be [$\frac{\Delta V}{\Delta(V-I)}$]$_{combined}=3.37\,\pm\,0.24$, which is similar to the value of [$\frac{\Delta V}{\Delta(V-I)}$]$_{galactic}=3.1\,\pm\,0.1$ calculated for galactic RCB stars (Skuljan et al. 2003). Extending this analysis by incorporating data obtained in other filters (e.g. $U$, $B$ and $R$) would enable the creation of a dust extinction curve, that would provide confirmation for the validity of the results obtained using the $V$ and $I$ data.

With only 77 RCB stars currently known, it is vital that this population, and the number of associated studies, be increased to enable the mysteries relating to their evolutionary status, dust formation mechanism and other aspects to be solved. Continuing the $UBVRI$ photometric monitoring programme with OGLE, MJUO and SALT, would be instrumental in narrowing the gap to find the answers to these questions. An increase in the amount of $UBVRI$ photometric data would permit the production of colour-magnitude diagrams ranging from $V$ vs ($U$-$B$) through to $V$ vs ($V$-$I$), therefore leading to quantification of the properties of circumstellar dust in the MC RCB stars.

The use of OGLE alerts to initiate more intensive photometric as well as spectorscopic campaigns on other telescopes (i.e. at MJUO and SALT), would broaden the scope of the study due to the availability of data collected over a wider wavelength range. In particular, obtaining spectroscopic observations during the initial decline phase, using the Robert Stobie Spectrograph on SALT, is another avenue that would be useful for enhancing the current RCB star knowledge.

The work undertaken for this paper may be thought of as the pilot project for the pursuit of additional research in this area. During the course of this research a number of hurdles have been encountered and overcome (e.g the method for data calibration), thus outlining a clear pathway along which significant future discoveries relating to the fascinating RCB phenonena can be made.

\section*{Acknowledgments} 

The data analysed for this paper were collected by several people at three different observatories around the world. First, we would like to acknowledge Pam Kilmartin, Alan Gilmore and Paul Tristram for making some of the observations at Mt John University Observatory during the 07/08 summer. Second, thanks to the Southern African Large Telescope astronomers, in particular Dr. Petri Vaisanen, for observing six of our candidate stars with SALT. SALT is a consortium consisting of the National Research Foundation of South Africa, Nicholas Copernicus Astronomical Center of the Polish Academy of Sciences, Hobby Eberly Telescope Founding Institutions, Rutgers University, Georg-August-Universitat Gottingen, University of Wisconsin-Madison, Carnegie Mellon University, University of Canterbury, United Kingdom SALT Consortium, University of North Carolina - Chapel Hill, Dartmouth College, American Museum of Natural History and the Inter-University Centre for Astronomy and Astrophysics, India. 

Thanks also to Simon Murphy for his initial investigation and subsequent generation of a list of possible RCB candidates. His research provided a firm foundation on which the work undertaken for this paper could be based. 

RMW would also like to acknowledge financial support received in the form of scholarships from the Department of Physics and Astronomy and the Dennis William Moore Fund. Thanks also needs to be extended to the Canterbury Branch of the Royal Society of New Zealand, the Dennis William Moore Fund, the Kingdon-Tomlinson Fund administered by the Royal Astronomical Society of New Zealand, the National Astronomical Research Institute of Thailand, and the International Astronomical Union, for their respective financial contributions that enabled RMW to participate in the Astronomical Society of Australia meeting (2007) and the Pacfic Rim Conference on Stellar Astrphysics (2008). Finally, AU would like to acknowledge the support for the OGLE project received in the form of the Polish MNiSW grant N20303032/4275.


\end{document}